\documentclass[12pt]{article}
\pdfoutput=1

\usepackage{amsmath}
\usepackage{amssymb}
\usepackage{epsfig}
\usepackage{graphicx}
\usepackage{cite}
\usepackage{multirow}
\usepackage{longtable}
\usepackage{lscape}
\usepackage{bbm}
\usepackage{dcolumn}
\usepackage{rotating}
\usepackage[squaren]{SIunits}

\allowdisplaybreaks[4] 

\newcommand{\capdef}{}
\newcommand{\mycaption}[2][\capdef]{\renewcommand{\capdef}{#2}%
        \caption[#1]{{\footnotesize #2}}}
\makeatletter
\renewcommand{\fnum@table}{\textbf{\tablename~\thetable}}
\renewcommand{\fnum@figure}{\textbf{\figurename~\thefigure}}
\makeatother

\newcounter{myenumi}

\renewcommand{\themyenumi}{\roman{myenumi}}
{\end{list}}

\newlength{\myem}
\newcounter{mysubequation}[equation]

\makeatletter
\renewcommand{\section}{\@startsection{section}{1}{0em}{-\baselineskip}%
{\baselineskip}{\normalfont\large\bfseries}}
\renewcommand{\subsection}%
{\@startsection{subsection}{2}{0em}{-0.7\baselineskip}%
{0.7\baselineskip}{\normalfont\bfseries}}
\makeatother

\newcommand{\bi}{\begin{itemize}}
\newcommand{\ei}{\end{itemize}}

\newcommand{\be}{\begin{equation}}
\newcommand{\ee}{\end{equation}}
\newcommand{\bea}{\begin{eqnarray}}
\newcommand{\eea}{\end{eqnarray}}

\newcommand{\ldm}{\Delta m_{31}^2}
\newcommand{\sdm}{\Delta m_{21}^2}
\newcommand{\deltacp}{\delta}
\newcommand{\stheta}{\sin^2 2 \theta_{13}}

\newcommand{\ie}{{\it i.e.}}

\newcommand{\eg}{{\it e.g.}}

\newcommand{\cf}{{\it cf.}}

\newcommand{\eq}{Eq.}

\newcommand{\fig}{Fig.}

\newcommand{\Ref}{Ref.}
\newcommand{\Refs}{Refs.}
\newcommand{\Sec}{Sec.}

\newcommand{\Tab}{Table}

\newcommand{\equ}[1]{\eq~(\ref{equ:#1})}
\newcommand{\figu}[1]{\fig~\ref{fig:#1}}

\begin{document}

\begin{titlepage}

\renewcommand{\thefootnote}{\alph{footnote}}

%


\renewcommand{\thefootnote}{\fnsymbol{footnote}}
\setcounter{footnote}{-1}

{\begin{center}
{\large\bf
Neutrino mass hierarchy determination with IceCube--PINGU 
} 

\end{center}}

\renewcommand{\thefootnote}{\alph{footnote}}

\vspace*{.3cm}
{\begin{center} {\large{\sc  
                Walter~Winter\footnote[1]{\makebox[1.cm]{Email:}
                winter@physik.uni-wuerzburg.de}
 		
                }}
\end{center}}
\vspace*{0cm}
{\it
\begin{center}

\footnotemark[1]
       Institut f{\"u}r Theoretische Physik und Astrophysik, \\ Universit{\"a}t W{\"u}rzburg, 
       97074 W{\"u}rzburg, Germany

\end{center}}

\vspace*{1cm}

\begin{center}
{\Large \today}
\end{center}

{\Large \bf
\begin{center} Abstract \end{center}  }

We discuss the neutrino mass hierarchy determination with atmospheric neutrinos in PINGU (``Precision IceCube Next Generation Upgrade''), based on a simulation with the GLoBES software including the full three flavor framework and parameter degeneracy, and we compare it to long-baseline experiment options. We demonstrate that the atmospheric mass hierarchy sensitivity depends on the achievable experiment properties and we identify the main targets for optimization, whereas the impact of a large number of tested systematical errors turns out to be small. Depending on the values of $\theta_{23}$, $\deltacp$, and the true hierarchy, a 90\%CL to $3\sigma$ discovery after three years of operation seems conceivable.
We also emphasize the synergy with existing beam and reactor experiments, driven by NO$\nu$A, such as the complementary coverage of the parameter space. Finally, we point out that a low intensity neutrino beam with a relatively short decay pipe could be used to determine the mass hierarchy with a sensitivity comparable to the LBNE experiment irrespective of the directional resolution of the detector.

\vspace*{.5cm}

\end{titlepage}

\newpage

\renewcommand{\thefootnote}{\arabic{footnote}}
\setcounter{footnote}{0}

\section{Introduction}

With the discovery of a non-zero value of $\theta_{13}$ by electron antineutrino disappearance reactor experiments~\cite{Abe:2011fz,An:2012eh,Ahn:2012nd}, the mixing angles and mass squared differences of the neutrinos are known at the percent level~\cite{GonzalezGarcia:2012sz,Fogli:2012ua,Tortola:2012te}; the flavor physics of the lepton sector has therefore become a precision discipline. The major outstanding issues in oscillation physics are the determination of the neutrino mass hierarchy ($\mathrm{sgn} ( \ldm)$) and the value of the leptonic Dirac CP phase $\delta$. If, in addition, $\theta_{23}$ is confirmed to be non-maximal, the $\theta_{23}$ octant needs to be determined. From \Ref~\cite{Huber:2009cw} it is clear that existing beam experiments, such as T2K~\cite{Itow:2001ee} and NO$\nu$A~\cite{Ayres:2004js} will most likely not allow for a high confidence level determination of any of these parameters -- even with the help of the reactor experiments and some mutual optimization of the neutrino-antineutrino running. Therefore, the next generation of experiments will be required. For the measurement of $\delta$, a new long-baseline neutrino oscillation experiment seems the straightforward choice because of excellent control over systematics; for a recent analysis and comparison, see \eg\  \Ref~\cite{Coloma:2012ji}.

The neutrino mass hierarchy is an excellent discriminator of flavor models~\cite{Albright:2006cw}, since (in the non-degenerate case) the structure of the effective mass matrix of the neutrinos strongly depends on $\mathrm{sgn} ( \ldm)$. The conventional technique to determine the hierarchy is using matter effects in long enough baselines~\cite{Wolfenstein:1978ue,Mikheev:1985gs,Mikheev:1986wj}. This approach can be used in a long-baseline neutrino oscillation experiment with a baseline $L \gtrsim 1000 \, \mathrm{km}$, such as a superbeam~\cite{Barger:2006vy,Barger:2007jq}, or using atmospheric neutrinos in an iron calorimeter~\cite{Samanta:2006sj,Ghosh:2012px,Blennow:2012gj} or a liquid argon detector~\cite{Barger:2012fx}. While a long-baseline neutrino oscillation experiment requires a new dedicated beam source, the atmospheric neutrino flux comes for free. The challenges are, however, completely different, as we will 
recover in this paper. Alternatives include
 the study of $\bar \nu_e$ disappearance over several oscillation peaks in a medium-baseline reactor experiment named Daya Bay-II; see, \eg\ \Ref~\cite{Li:2013zyd}. This technique is different from using matter effects, and requires excellent energy resolution. Furthermore, cosmological tests of the mass hierarchy can be performed, see \eg\ \Ref~\cite{Oyama:2012tq}.

A new way to use atmospheric neutrinos to determine the neutrino mass hierarchy has recently been drawing some attention: increasing the Digital Optical Module (DOM) density in a neutrino telescope, operated in sea water or ice, lowers the threshold and allows for a Megaton-size detector already at 10~GeV. Such an approach has been identified for IceCube-DeepCore at the South Pole, called PINGU (``Precision IceCube Next Generation Upgrade'')~\cite{Koskinen:2011zz,Clark:2012hya}, and ANTARES/KM3NeT in the Mediterranean, called ORCA (``Oscillation Research with Cosmics in the Abyss'')~\cite{ORCA}. 
The respective physics potential of these ideas has been studied in \Refs~\cite{Akhmedov:2012ah,Agarwalla:2012uj,Franco:2013in,Ohlsson:2013epa,Esmaili:2013fva,Ribordy:2013xea}. We focus on the neutrino mass hierarchy determination in PINGU in this paper. The main selling points of this idea are the relatively short timescale compared to any other option which requires the excavation a large underground cavern, the very predictable and moderate cost, and the small risk due to the experience with the existing technology.  

We first of all discuss the physics potential using atmospheric neutrinos including the full three flavor framework, oscillation parameter correlations and degeneracies, and an unprecedentedly large number of systematical errors. We discuss the implementation of PINGU in \Sec~\ref{sec:implementation}, the main impact factors including systematics in \Sec~\ref{sec:impacts}, and the performance in \Sec~\ref{sec:performance}. Then in \Sec~\ref{sec:existing}, we compare the potential of PINGU to the one of the existing beam and reactor experiments. Finally, we point out that one may also target a low intensity neutrino beam from one of the major high-energy physics laboratories on the Northern hemisphere towards the South Pole in \Sec~\ref{sec:beam}. Earlier studies discussing this possibility have been, for instance,  \Refs~\cite{Dick:2000fn,Winter:2005we,Fargion:2010vb,Tang:2011wn}.

\section{PINGU implementation and systematics}
\label{sec:implementation}

For the simulation of PINGU, a modified version of the GLoBES (``General Long Baseline Experiment Simulator'') software~\cite{Huber:2004ka,Huber:2007ji} is used.\footnote{This version is based on a (yet) unpublished prototype for GLoBES 4.0, which was originally written for \Ref~\cite{Coloma:2012ji}. The advantage of this prototype is that the different directional bins can be defined as different experiments, and systematics can be easily and self-consistently defined using the pull method. A more detailed description can be found in \Ref~\cite{Coloma:2012ji}.} The directional smearing is performed after the event rate computation, whereas the energy resolution is a built-in feature of GLoBES. Note that the error 
by introducing the directional smearing after the event rate distribution
can be shown to be second order in $E' - E$ if the 
directional resolution depends on the incident energy $E'$ instead of the reconstructed neutrino energy $E$.
The energy binning is chosen from $2$ to $50 \, \mathrm{GeV}$ in steps of 1~GeV.
The directional binning is chosen from $\cos \theta_z=-1$ to $0$ with equidistant steps $\Delta \cos \theta_z = 0.05$ to follow \Ref~\cite{Akhmedov:2012ah}. We take into account the background of down-going neutrinos which may be re-constructed as up-going events (effective $4\pi$ coverage) by defining two ``overflow'' bins $0 \le \cos \theta_z \le 0.2$ and $0.2 \le \cos \theta_z \le 1.0$. Note that the down-going neutrinos create a non-oscillating background for the mass hierarchy determination, while the overflow bins can be used to constrain systematical errors. In addition, note that the binning has been checked to be sufficiently fine-grained not to affect the sensitivity significantly, whereas in principle larger bin widths are allowed for quasi-horizontal events. At low energies, where the oscillation probability changes very quickly, the sampling density has been increased to 50~MeV intervals to avoid aliasing effects.

The directional smearing (redistribution of events) between incident $\theta_z'$ and reconstructed $\theta_z$ is computed by a (normalized) Gaussian
\begin{equation}
 R(\theta_z,\theta_z',E) = \frac{1}{\Delta \theta_z(E)  \sqrt{2 \pi}} \exp \left( - \frac{(\theta_z-\theta_z')^2}{2 (\Delta \theta_z(E))^2} \right) \, . \label{equ:gmap}
\end{equation}
with the resolution $\Delta \theta_z(E)$. Since the directional resolution depends on energy, the fraction of events with incident direction $\theta_z'$ mapped into bin $i$ with boundaries $(\cos \theta_z)_{i,\mathrm{min}}$ and  $(\cos \theta_z)_{i,\mathrm{max}}$ has been pre-computed by
\begin{equation}
 f_i(\theta_z',E)  = \int\limits_{(\theta_z)_{i,\mathrm{min}}}^{(\theta_z)_{i,\mathrm{max}}} R(\theta_z,\theta_z',E) d\theta_z \, ,
\end{equation}
which is also often called a ``migration matrix'', since $\theta_z$ and $E$ can only take fixed values in the simulation. As additional complication, it has to be taken into account that up- and downgoing events are properly mapped into the bins (in \equ{gmap}, $\theta_z$ may take values smaller than zero or larger than $\pi$, which just corresponds to shifting the azimuth by $\pi$ and correcting the angle). The migration matrices have been directly compiled into the GLoBES code.

For the atmospheric neutrino flux, we use the most up-to-date South Pole flux~\cite{Athar:2012it}, where we use the azimuth-averaged solar-min version. In order to take into account site- and model-dependent uncertainties, we include a substantial normalization error of the order 20\%, see \Ref~\cite{Franco:2013in}, a slope error (``zenith bias'') of the order of 4\%~\cite{Honda:2006qj} as implemented in \Ref~\cite{Esmaili:2013fva} on the up-going (signal) events, an (uncorrelated) error on the down-going event sample of the same order, and errors on flavor ratios and polarities estimated from \Ref~\cite{Honda:2006qj}. Our PINGU (default) fiducial mass is based on the official version 12/2012 using a minimum of 20 hits/event, also known as V6.%
\footnote{The fiducial mass is linearly interpolated among $(E \, [\mathrm{GeV}],M_{\mathrm{fid}} \, [\mathrm{Mt}]) =(1,0)$, $(4,1.5)$, $(6,2.4)$, $(8,2.9)$, $(10,3.3)$, $(12,3.5)$, $(26,4.3)$, and $(50,4.9)$~\cite{PINGU}.} 
It includes 20 strings with a $26 \, \mathrm{m}$ string-string spacing, and 60 DOMs per string with a $5 \, \mathrm{m}$ DOM-DOM spacing. It should be noted that the actual PINGU configuration is still subject to optimization, and that smaller or larger fiducial masses and threshold functions are conceivable depending on the number of strings and DOM density. For instance, a 40 string setup may perform closer to the ``optimistic'' setup, we define below.

We use three years of exposure, unless noted otherwise. The event rates have been cross checked with \Ref~\cite{Akhmedov:2012ah,Agarwalla:2012uj,Ohlsson:2013epa}. 
The oscillation parameters and their uncertainties are taken from \Ref\cite{GonzalezGarcia:2012sz}, where it is noteworthy that two solutions with $\theta_{23} \simeq 40^\circ$ and $50^\circ$ exist. For the true $|\ldm|$, we choose the normal hierarchy best-fit $2.473 \, 10^{-3} \, \mathrm{eV}^2$. We do not take into account the external errors on $\ldm$ and $\theta_{23}$ though to avoid a bias from the external fit or a preference of a particular octant. In some cases, we combine the PINGU data with the equivalent of 10 years of disappearance data from T2K, based on the simulation in \Ref~\cite{Huber:2009cw}, which is expected to provide the best available knowledge at least on $\ldm$ at the analysis time -- a method which has proven to be successful for beta beam simulations~\cite{Huber:2005jk,Winter:2008cn}. We have checked that the T2K disappearance data alone do not have any significant mass hierarchy sensitivity.

\begin{table}[t!]
\begin{center}
\begin{tabular}{p{3.7cm}rrrp{5cm}c}
\hline
  Systematics & Opt. & Def. & Cons. & Comments & Ref. \\
\hline
 \multicolumn{6}{l}{\bf Experiment properties:} \\
Fiducial mass and \newline energy threshold & (*) & 12/2012 & 12/2012 & (*) Improved threshold from denser instrumentation, starting at $2 \, \mathrm{Mton}$ at $1 \, \mathrm{GeV}$ linearly increasing to $2.6 \, \mathrm{Mton}$ at $6 \, \mathrm{GeV}$ & \cite{PINGU} \\
Energy res. $\Delta E/E$  & 0.15 & 0.25 & 0.35 & &  \\
Dir. resolution $\Delta \theta_z$ & $0.5 \sqrt{\frac{m_p}{E}}$ & $0.6 \sqrt{ \frac{m_p}{E} } $ & $\sqrt{\frac{m_p}{E}}$ & ``Default'' corresponds to $10^\circ$ at $10 \, \mathrm{GeV}$ & \cite{Akhmedov:2012ah} \\
Cascade mis-ID frac. & 0.01 & 0.05 & 0.1 \\[0.2cm]
 \multicolumn{6}{l}{\bf Systematical uncertainties:} \\
Normalization & 0.10 & 0.25 & 0.35 & Includes fiducial mass, atmospheric flux models \\
mis-ID cascades & 0.05 & 0.075 & 0.10 & Uncorrelated among electromagnetic, hadronic, and NC cascades & \\
Cross sections (DIS) & 0.05 & 0.075 & 0.10 & Uncorrelated between neutrinos and antineutrinos 
& \cite{Coloma:2012ji} \\
Matter density & 0.005 & 0.01 & 0.05 & Uncorrelated among all directional bins & \cite{Geller:2001ix}
\\[0.2cm]
 \multicolumn{6}{l}{\bf Uncertainties of atmospheric neutrino flux:} \\
Normalization & \multicolumn{4}{l}{Included in ``Normalization'' above.} & \cite{Franco:2013in} \\
Slope error \newline (zenith bias) & 0.01 & 0.04 & 0.10 &  & \cite{Esmaili:2013fva,Honda:2006qj} \\
Flavor $\nu_e$/$\nu_\mu$ & 0.005 & 0.01 & 0.02 & & \cite{Honda:2006qj} \\
Polarity $\bar \nu_\mu$/$\nu_\mu$ & 0.01 & 0.02 & 0.03 & & \cite{Honda:2006qj} \\
Polarity $\bar \nu_e$/$\nu_e$ & 0.01 & 0.025 & 0.03 & & \cite{Honda:2006qj} \\
BG down-going & 0.05 & 0.075 & 0.10 & & \cite{Honda:2006qj} \\
\hline
\end{tabular}
\end{center}
\mycaption{\label{tab:sys} Considered systematical errors and experiment properties in this analysis, where the assumptions for three different scenarios (optimistic, default, conservative) are listed.  }
\end{table}

The considered systematical errors and experiment properties considered in this analysis are listed in  \Tab~\ref{tab:sys}. In order to the test impact of systematics, we have defined three values in each case (optimistic, default, conservative), which are supposed to represent the performance of the experiment in the best, conceivable, and worst case. Apart from the systematics already discussed above, an improved threshold has been considered for the optimistic setup, as it may be achieved by a denser instrumentation, such as a 40 string configuration. The energy and directional resolution corresponds to similar assumptions as used in the literature~\cite{Akhmedov:2012ah}, where the default values correspond to $\Delta E=2.5 \, \mathrm{GeV}$ and $\Delta \theta_z=10^\circ$ at $10 \, \mathrm{GeV}$ (where the main significance comes from, as we will show below). The fraction of cascades mis-identified as muon tracks is, at this point, an educated guess, which may actually be a function of energy. For the cross sections and matter density uncertainty, the assumptions from \Ref~\cite{Coloma:2012ji} have been adopted.
 We will, however, demonstrate below that only very few of these systematics are important for the analysis at all. We have also tested an energy calibration error (for the definition, see GLoBES manual~\cite{Huber:2004ka}), and we have not found any significant impact, even not on the $\ldm$ measurement. The reason is that the solar $\sdm$ is well constrained externally, which means that the bins with long baselines and low energies can be used for the energy calibration (the energy calibration error would affect both the solar and atmospheric oscillations).
Since the inclusion of the energy calibration is computationally expensive, we have not included it in the main line of this work.

\section{Performance indicator, main impacts, and potential for optimization}
\label{sec:impacts}

\begin{figure}[t!]
\begin{center}
 \includegraphics[width=0.49 \textwidth]{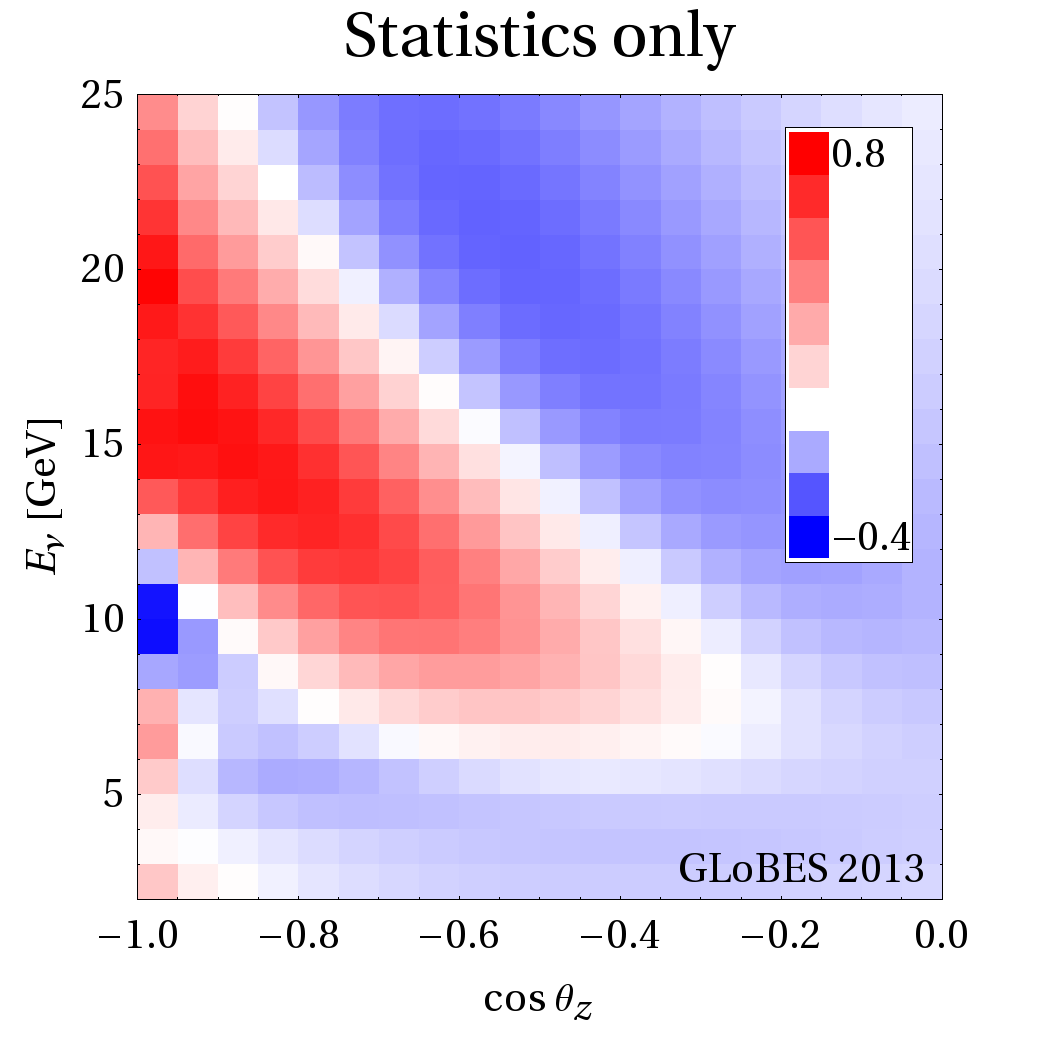}
 \includegraphics[width=0.49 \textwidth]{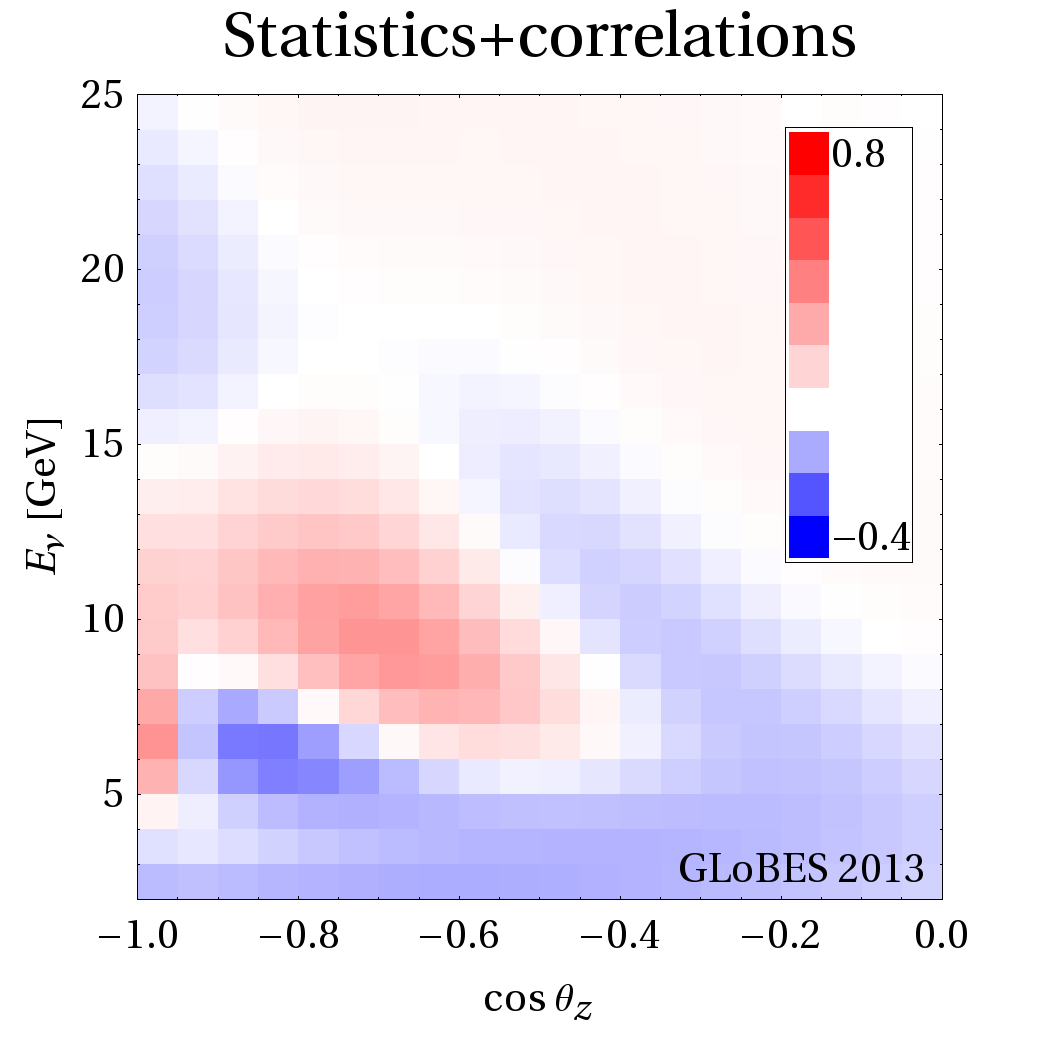}
\end{center}
\mycaption{\label{fig:statcorr} Statistical significance $(N_\mu^{\mathrm{IH}}-N_\mu^{\mathrm{NH}})/\sqrt{N_\mu^{\mathrm{NH}}}$ per bin for the mass hierarchy determination in PINGU (default setup, $\theta_ {23}=40^\circ$, $\deltacp=0$). The left panel includes statistics only ($\ldm \rightarrow -\ldm$), the right panel uses the IH rates at the fit minimum, \ie, it includes the oscillation parameter correlations. The minimum in this case is found at 
$\theta_{12}=0.5824$, $\theta_{13}=0.1526$, $\theta_{23}=0.6987$,  $\deltacp=2.5398$, $\sdm=7.499 \, 10^{-5} \, \mathrm{eV}^2$, $\ldm=- 2.337 \,10^{-3} \mathrm{eV}^2$. The total statistical significance, as defined here using the Gaussian $\chi^2$, is $66$ and $4.2$ in the left and right panels, respectively. This calculation has been performed without systematical errors.
}
\end{figure}

We define the sensitivity to the hierarchy as the minimal $\chi^2$ of all solutions with the flipped hierarchy, compared to the simulated hierarchy where the minimal $\chi^2=0$. Therefore, oscillation parameter correlations (connected degenerate solutions) and degeneracies (such as in the $\theta_{23}$ octant) are automatically included if all possible degenerate solutions are properly located. Note that this way to define the hierarchy sensitivity is consistent with the literature on long-baseline neutrino oscillation experiments, and can therefore be used for a direct comparison. Occasionally we will show  the number of sigmas ($N\sigma$) for the discovery assuming that $N \sigma = \sqrt{\chi^2}$ for 1 d.o.f. Note that this frequently used assumption may not exactly apply to the hierarchy sensitivity, see \Ref~\cite{Ciuffoli:2013rza} and references therein for a more detailed discussion. However, we use it equally for all experiments.

We show the impact of the correlations in  \figu{statcorr}. 
First of all, consider the statistical significance per bin in the left panel, which is to be compared to Fig.~13 in \Ref~\cite{Akhmedov:2012ah}.\footnote{In this figure, we use a Gaussian $\chi^2$, whereas for the main analysis the Poissonian $\chi^2$ of GLoBES is used.} Apart from the different fiducial mass as a function of energy and slightly different angular and directional resolutions, the result is very similar. If, however, the oscillation parameters of the fit solution are allowed to take any value in the parameter space region $\ldm <0$ (are minimized over), the significance decreases, see right panel; the actual position of the minimum is given in the figure caption. 
In fact, after the marginalization, the main significance comes from mantle-crossing baselines ($L \sim 6000 - 9000 \, \mathrm{km}$) at about 9~GeV (dark red region). This figure has to be interpreted with care, though, since it only shows where the resulting tension between the two hierarchies remains -- whereas other regions help to reduce systematical uncertainties and the measurements of the oscillation parameters.\footnote{For instance, we have not shown the energy range between 25 and 50~GeV in the figures at all, where the contribution to the total statistical significance for the mass hierarchy is small in the right panel. However, without this region, the minimal $\chi^2$ decreases because it helps to measure some of the oscillation parameters and systematical errors.}

The most important correlation comes from the definition of $\ldm \equiv m_3^2 - m_1^2$. For the normal hierarchy, it is the difference between the heaviest and lightest (squared) masses; for the inverted hierarchy, it is the difference between the lightest and second-heaviest (squared) masses. Therefore, the fit values for the normal and inverted hierarchies are different as an artefact of an asymmetric definition of $\ldm$, and the flipped hierarchy fit is not exactly found at $\ldm \rightarrow - \ldm$. A possible way out is to define a symmetrized  $(\ldm)_{\mathrm{eff}}$, which reads for the $\nu_\mu$ disappearance channels~\cite{deGouvea:2005hk,Nunokawa:2005nx,deGouvea:2005mi,Ghosh:2012px,Agarwalla:2012uj,Blennow:2012gj}
\begin{equation}
  (\ldm)_{\mathrm{eff}} = \ldm - \sdm (\cos^2 \theta_{12} - \cos \deltacp \, \sin \theta_{13} \, \sin 2 \theta_{12} \, \tan \theta_{23}) \, . \label{equ:sym}
\end{equation}
 However, note that it contains other oscillation parameters, such as $\deltacp$, which itself can vary. 
 Therefore, we choose complete minimization over all fundamental parameters as method for the hierarchy determination within the full three flavor oscillation framework, as opposed to \equ{sym}. As this is a built-in feature of GLoBES, it is automatically included as for any other long-baseline experiment.

\begin{figure}[t!]
\begin{center}
 \includegraphics[width=\textwidth]{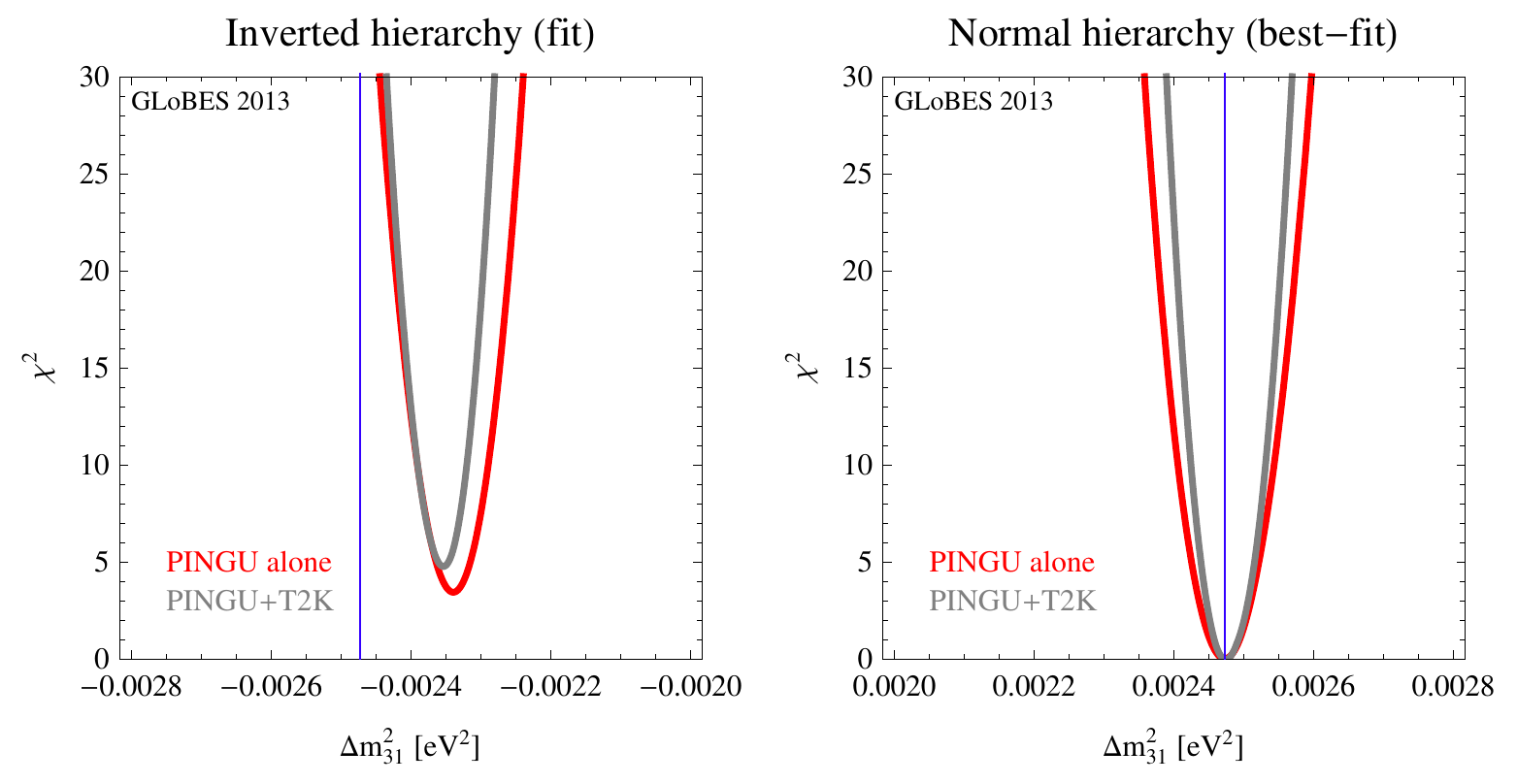}
\end{center}
\mycaption{\label{fig:dmall} The $\chi^2$ as a function of $\ldm$ for a simulated normal hierarchy (and $\deltacp=0$, $\theta_{23}=40^\circ$). The left panel shows the inverted hierarchy fit, the right panel the best-fit region. The different curves correspond to PINGU alone, and PINGU combined with T2K disappearance data. Vertical lines correspond to $\pm \ldm$, where $\ldm$ is the simulated best-fit value.  Default setup from \Tab~\ref{tab:sys} used.
}
\end{figure}

We illustrate the $\ldm$ correlation in \figu{dmall}, where the $\chi^2$ as a function of $\ldm$ is shown for the best-fit (right panel) and degenerate solution (left panel), see red curves. The minimal $\chi^2$ in the left panel corresponds to the hierarchy sensitivity, whereas the minimal $\chi^2$ in the right panel is zero by definition (statistical fluctuations are not simulated here, we only simulate the ``average'' experiment). If the fit $\ldm$ is fixed to $-|\ldm|$, as illustrated by the vertical line in the left panel, a large artificial $\chi^2$ is introduced because PINGU itself has a good $\ldm$ precision. Similarly,  such a contribution can arise from a tension of different data, such as external knowledge on $\ldm$ added at some position, which is {\em not} to be interpreted as actual mass hierarchy sensitivity. 

It is also an interesting question if such external knowledge on $\ldm$ can actually help; we therefore include the T2K disappearance data in the gray curves.  The precision PINGU can achieve alone for $\ldm$ is, for our simulation, about 0.9\%, compared to  the current precision of about 3\%~\cite{GonzalezGarcia:2012sz} and the precision of the T2K data of about 1.3\%, which means that the inclusion of external data may not be necessary. The combined precision of PINGU and T2K, which can be read off from the figure, is about 0.7\%. However, 
the addition of the T2K data  only marginally affects the position of the minimum or the minimal $\chi^2$ -- and therefore only marginally affects the hierarchy sensitivity. We therefore do not include it in the following sections, unless noted otherwise. It is also noteworthy that the position of the minima in \figu{dmall}, left panel, are not exactly matching, which means that even \equ{sym} can only provide a rough estimate if the precision on $\ldm$ is as high as in these experiments. For instance, computing the hierarchy sensitivity for a fixed fit $\ldm$ obtained from \equ{sym} yields a $\chi^2 \simeq 8.8$, which is larger than the actual mass hierarchy sensitivity at the minimum. We therefore rely on full minimization, and do not discuss the effect of $\ldm$ separately.

\begin{figure}[t!]
\begin{center}
 \includegraphics[width=\textwidth]{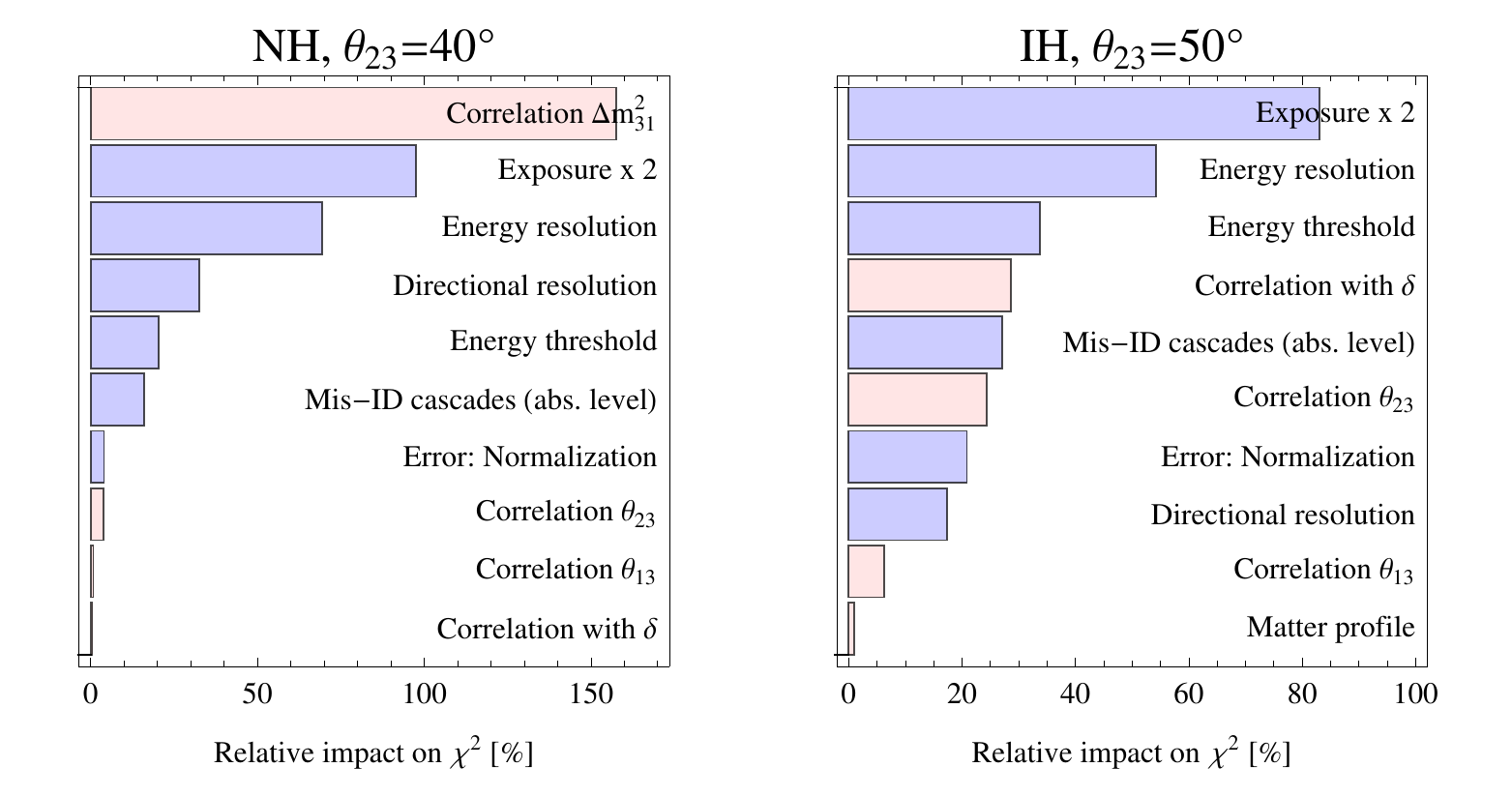}  
\end{center}
\mycaption{\label{fig:sysres} Dominant impact factors (as relative effect on the total $\chi^2$) for two different parameter sets (see plot labels) and  $\deltacp=0$.  Impact factors under experiment control are shown in blue/dark. 
}
\end{figure}

In order to address the potential for optimization, we need to identify the main impact factors for the sensitivity. For that, we have tested all the properties listed in \Tab~\ref{tab:sys} including the total exposure (running time $\times$ detector mass) on the default systematics setup, one by one, for two different sets of parameters. In addition, we have tested the impact of parameter correlations. As method,
systematics are switched off to test their impact, where the resulting (relative) change of the total $\chi^2$ is computed.
Correlations are checked by evaluting the full parameter degeneracy and then fixing only one oscillation parameter at a time.
 Threshold, energy resolution, and directional resolution are switched to the optimistic values in \Tab~\ref{tab:sys} and mis-identified cascades are switched off to test their effect. In addition, the impact of a factor two higher exposure (running time or fiducial mass) is tested.
Although these choices are to some degree arbitrary, they help to identify the relevant systematics and impacts.
This method dates back to \Ref~\cite{Huber:2002mx} and has been motivated by the idea to identify the impacts with the greatest potential for optimization. It typically leads to the same qualitative results as switching all systematics off and then each of them on, one by one; different methods have been explicitely tested in \Ref~\cite{Coloma:2012ji}. However, it should be noted that it is not suitable to quantitatively infer the combined effect of different impact factors, as we will show later for the optimistic setup, since, for instance, systematics can be correlated with each other or with oscillation parameters. It rather illustrates what happens if a single of the main impacts can be improved. 

The impact on the total $\chi^2$ is shown in \figu{sysres} for the ten most dominant impact factors for two different sets of parameters, \ie, all of the shown impacts are important. It is first of all interesting to observe that the mass hierarchy determination is rather insensitive to systematical errors, such as on cross sections or on the atmospheric neutrino flux, which can be determined by a large number of bins without mass hierarchy sensitivity; \cf, \figu{statcorr}.  The main impacts, which may be optimized for, are represented by the blue (dark) bars:
\begin{description}
 \item[Exposure/fiducial mass] May be improved by adding an event sample with un-contained muon tracks, by allowing for less that 20 hits for the threshold, by adding cascade data, or by increasing the instrumentation volume.
 \item[Directional resolution] May be improved by a denser instrumentation.
 \item[Energy resolution] May be improved by better analysis techniques or better DOM hardware.
 \item[Energy threshold] Can be improved by a denser instrumentation.
  \item[Mis-ID cascades] Biggest unknown. Has to be addressed,  since it is important.
 \item[Normalization] Requires both control over fiducial mass of the experiment and the atmospheric neutrino flux prediction.
\end{description}
In addition, note that several of these properties may be improved by using inelasticity information, which might also allow for separate neutrino and antineutrino samples~\cite{Ribordy:2013xea}. While some improvements may rely on more advanced analysis techniques to be developed and improved within the next years, some require a modification of the experiment design beforehand. Note that the order of the impacts on \figu{sysres} in many cases depend on the difference between the values for the default and optimistic setups in \Tab~\ref{tab:sys}, which means that all of the listed factors are important. In addition, our qualitative conclusions are rather independent of the parameter space region, as it can be inferred from the comparison between the left and right panels of \figu{sysres}.

As far as correlations are concerned (red/light bars), their individual impact depends on how well the location of the degenerate solution is estimated. Here we use \equ{sym} to ``guess'' the degenerate $\ldm$, and use the simulated values for the other parameters. For instance, in the left panel, \equ{sym} does not seem to work well enough and the correlation with $\ldm$ is the first one to be taken into account. In the right panel, \equ{sym} works better, but the degeneracy is located at a different  $\deltacp$, $\theta_{23}$, and $\theta_{13}$ than the simulated values. This exercise illustrates that the full parameter degeneracy has to be taken into account, since whatever  estimator for the degenerate solution is chosen (here \equ{sym}), it does not work equally well everywhere in parameter space.  In the following, we therefore take into account the full parameter degeneracy everywhere, \ie, we take into account the actual parameter space topology and do not rely on such estimators.

\section{Performance using atmospheric neutrinos}
\label{sec:performance}

\begin{figure}[p!]
\begin{center}
 \includegraphics[width=\textwidth]{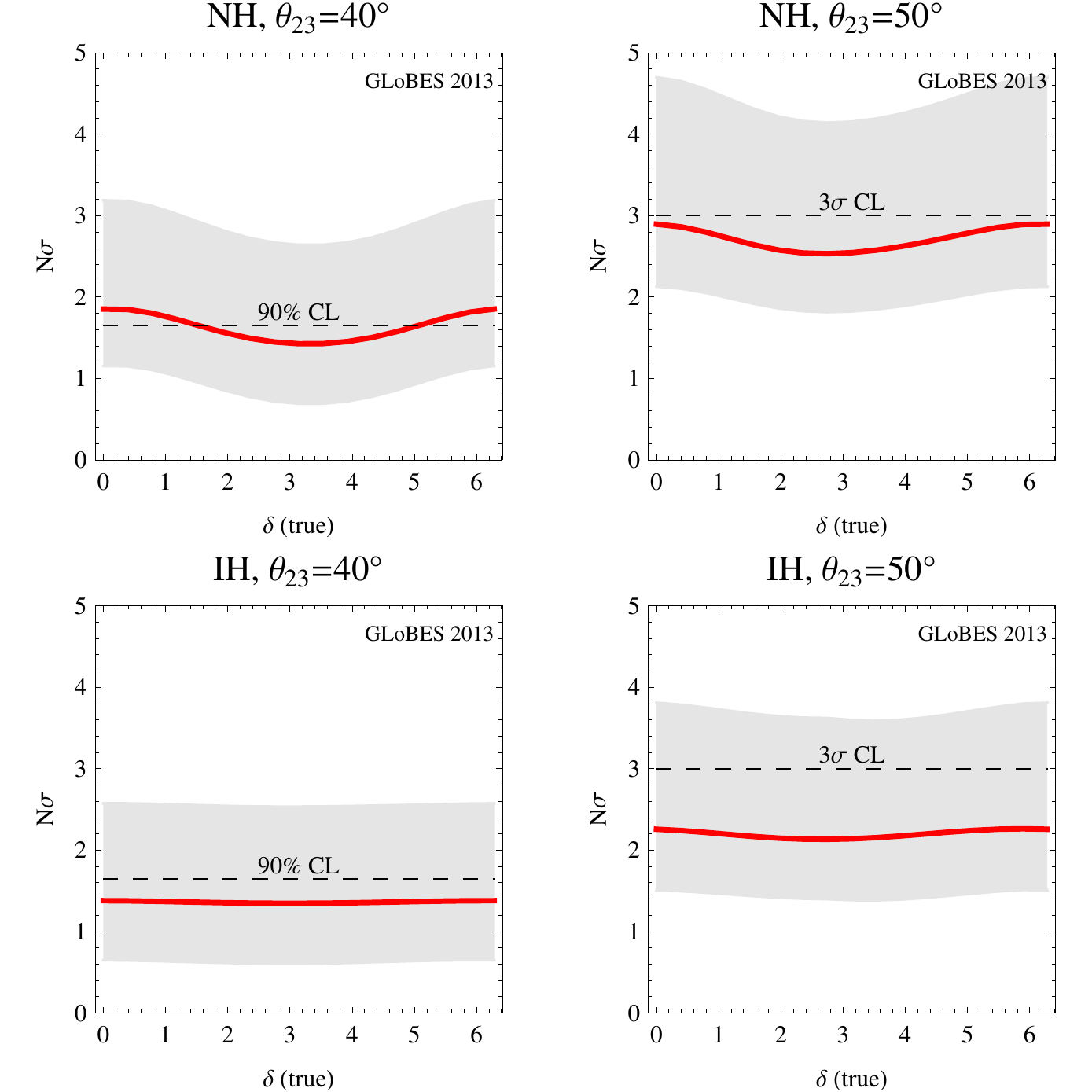}
\end{center}
\mycaption{\label{fig:sensall} The number of sigma (N$\sigma$) for the mass hierarchy discovery as a function of the true $\deltacp$ for the different (true) hierarchies (rows) and octants (columns), as given in the plot captions,  for three years of data taking. The solid curves correspond to the default setup in \Tab~\ref{tab:sys}, the shaded region shows the impact of systematics between optimistic (upper end) and conservative (lower end). 
}
\end{figure}

The performance of PINGU is shown in \figu{sensall} as a function of $\deltacp$ for both true hierarchies and both octant $\theta_{23}$ solutions. The solid curves correspond to the default setup in \Tab~\ref{tab:sys}, the shaded region shows the impact of systematics between optimistic (upper end) and conservative (lower end). From the figure it is clear that PINGU performs significantly better if $\theta_{23}= 50^\circ$ rather than $40^\circ$, and if the true hierarchy is normal rather than inverted.\footnote{The sensitivities are somewhat worse for the inverted hierarchy, since the event rates in the $\nu_\mu$ appearance channels will be somewhat lower because of the lower cross sections. }

\begin{figure}[t!]
\begin{center}
 \includegraphics[width=0.46\textwidth]{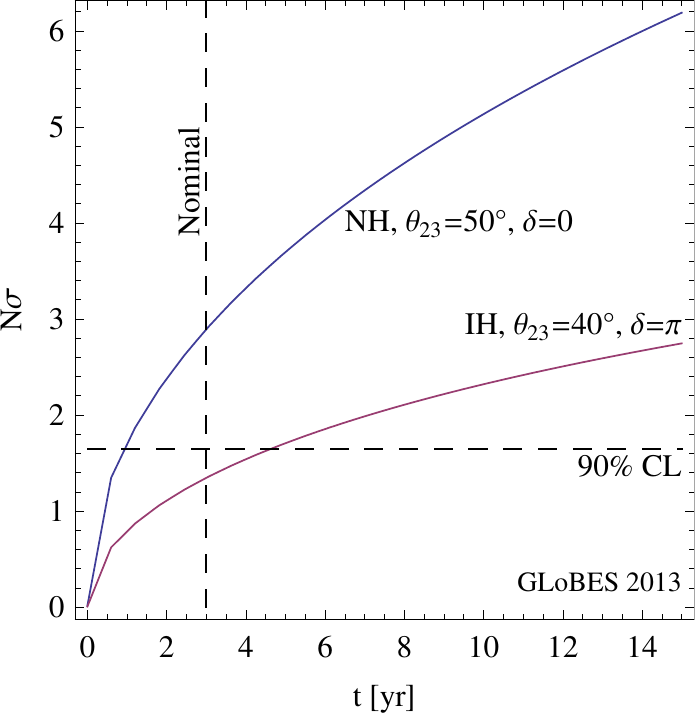} \hspace*{0.03\textwidth}
  \includegraphics[width=0.445\textwidth]{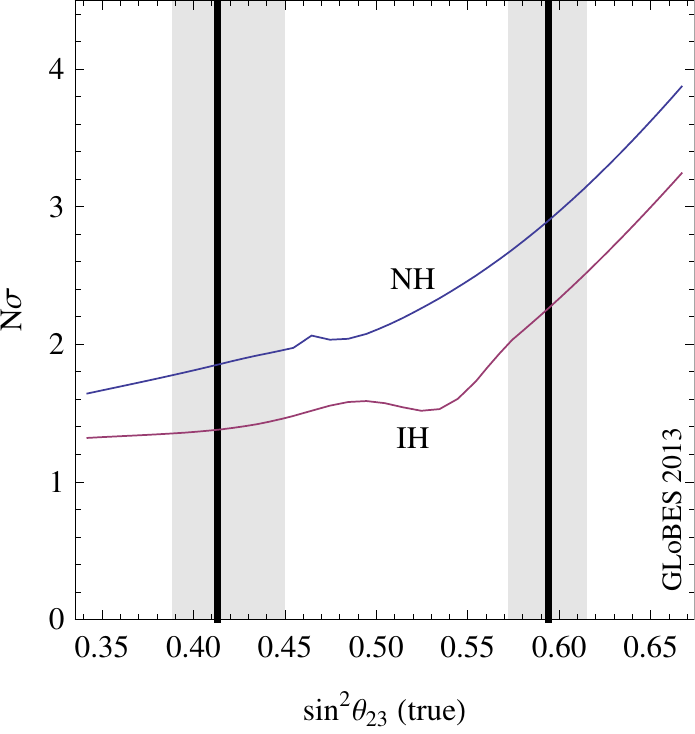} 
\end{center}
\mycaption{\label{fig:expplot} Left panel: Number of $\sigma$ for the hierarchy discovery as a function of time for two extreme cases of the true parameter values (see legend). Right panel: Number of $\sigma$ for the hierarchy sensitivity as a function of $\sin^2 \theta_{23}$ (true) for the different hierarchies. The plot range corresponds to the current $3\sigma$ allowed range for $\sin^2 \theta_{23}$, and the best-fit values and $1 \sigma$ ranges are marked by vertical lines and shadings, respectively~\cite{GonzalezGarcia:2012sz}. The default setup has been used in both panels (three years of data taking and $\deltacp=0$ in right panel).
}
\end{figure}

For $\theta_{23}=50^\circ$, a 90\% CL hint can be reached after three years even under conservative assumptions, and $3 \sigma$ are conceivable for a somewhat improved setup (compared to the default setup) and a normal hierarchy. For $\theta_{23}= 40^\circ$, however, considerably larger exposures will be needed for a $3\sigma$ discovery. For the default setup, a 90\% CL hint after three years seems conceivable in almost all cases (solid curves). It is however clear that the final performance will depend on the true hierarchy, $\deltacp$, and $\theta_{23}$, apart from the detector performance. In turn, one has to adjust the running time  to accumulate the necessary statistics.  This is, for the most extreme cases, illustrated in \figu{expplot} (left panel) for the default systematics. In one case, a $2\sigma$ hint can be reached after one year and almost $3\sigma$ evidence after three years, in the other case, the 90\% CL hint can be reached after five years, and $3 \sigma$ are out of reach. Note that also $\theta_{13}$ within the currently allowed $3 \sigma$ range can impact the performance~\cite{Ghosh:2012px,Blennow:2012gj}. However, we do not treat $\theta_{13}$ at the same level as the other unknown parameters, since the precision of $\theta_{13}$ is continuously being improved; on the other hand, it cannot be anticipated that the other discussed parameters (hierarchy, $\deltacp$, $\theta_{23}$ octant) will be known before the construction or even analysis time~\cite{Huber:2009cw}.

While the two best-fit solutions for $\theta_{23}$ in \Ref~\cite{GonzalezGarcia:2012sz} favor non-maximal atmospheric mixing, the $3 \sigma$ allowed range is much larger and includes the possibility of maximal atmospheric mixing. We therefore show in \figu{expplot}, right panel, the $\theta_{23}$ dependence in the currently allowed $3 \sigma$ range. Obviously, the sensitivity for the normal (inverted) hierarchy can vary between $1.6 \sigma$ and $3.9 \sigma$ ($1.3 \sigma$ and $3.3 \sigma$) for three years of data taking within the currently allowed $3 \sigma$ range, depending on the actual value of $\sin^2 \theta_{23}$. The kinks in this plot come from the octant degeneracy, \ie, the minimal $\chi^2$ is found in the inverted hierarchy, inverted octant solution, and may be eliminated if the octant was known at the analysis time. A definitive conclusion on the $\theta_{23}$ octant from existing equipment is, however, unlikely~\cite{Huber:2009cw} (see Fig.~5 therein). 

\section{Comparison with existing beam and reactor experiments}
\label{sec:existing}

Compared to PINGU, other experiments will be sensitive to the mass hierarchy at the same timescale, where we especially focus on existing equipment with possible upgrades. The most sensitive experiment is the long-baseline neutrino oscillation experiment NO$\nu$A in the US, with a baseline of 810~km. In addition, data from the reactor experiments (Double Chooz, Daya Bay, and RENO) will be available, as well as data from T2K. In \Ref~\cite{Huber:2009cw}, the time evolution for the sensitivities has been discussed as a function of the exposure of the individual experiments. We adopt these assumptions for the time evolution, and re-evaluate the sensitivities for the current best-fit parameters.

We consider to scenarios, called ``2020'' and ``2025'', which are identical to \Ref~\cite{Huber:2009cw}:
\begin{description}
\item[2020]
Beam and reactor experiments: The main sensitivity to the mass hierarchy comes from NO$\nu$A, which is assumed to start 08/2012 with full beam (0.7~MW) but 2.5~kt detector mass only, linearly increasing to 15~kt until 01/2014. This running plan is, in fact, to so far away from the current status, the final luminosity is expected to be reached 05/2014, see \Ref~\cite{BackhouseNOVA}. There are also assumptions for T2K and the reactor experiments, the final sensitivity is however not so sensitive to. PINGU: Since PINGU is expected to be completed in Feb. 2017 or 2018, we compare this sensitivity to three years of PINGU data.
\item[2025] Beams and reactor experiments: Here it is assumed that the beams are upgraded. In particular, ProjectX for NO$\nu$A is assumed to an increase of the beam power from 0.7 to 2.3 MW from March 2018 to March 2019. PINGU: We compare this sensitivity to eight years of PINGU data.
\end{description}

\begin{figure}[t]
\begin{center}
 \includegraphics[width=0.48\textwidth]{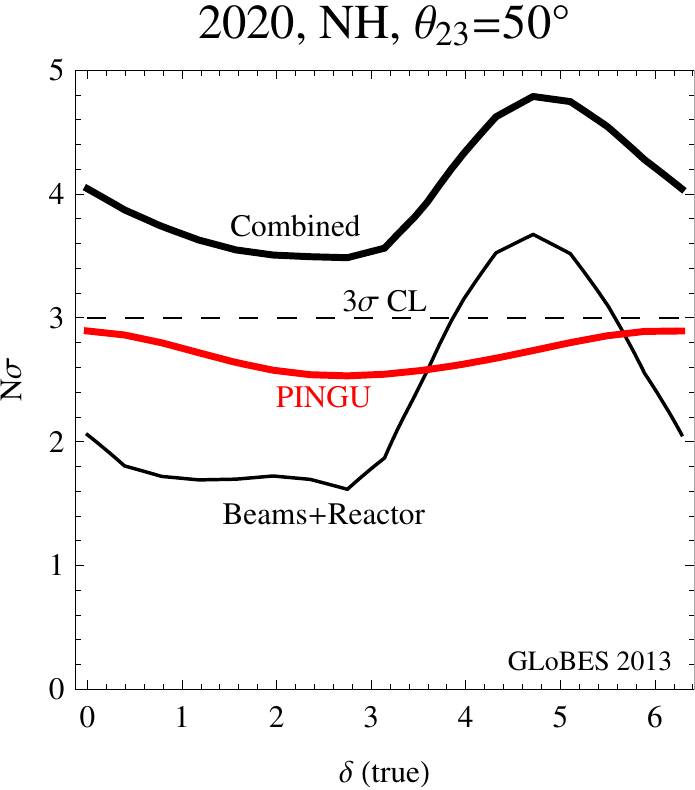} \hspace*{0.02\textwidth} %
 \includegraphics[width=0.48\textwidth]{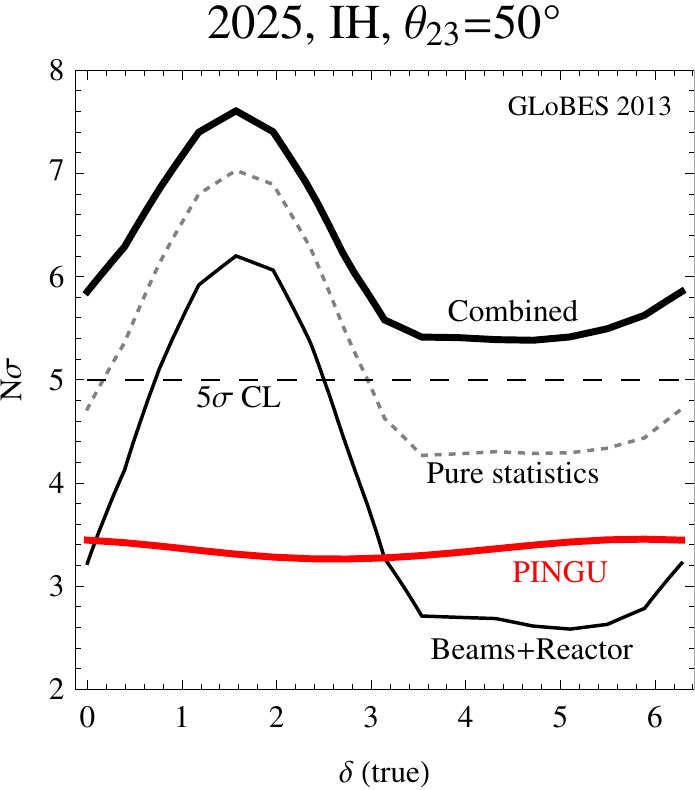}
\end{center}
\mycaption{\label{fig:synergy} Number of $\sigma$ for the hierarchy discovery as a function of $\deltacp$ for two different scenarios for PINGU (three years, left panel and eight years, right panel, respectively), beams and reactor experiments (scenario 2020, left panel, and 2025, right panel, respectively), and their combination. In the right panel, the hypothetical pure combination with the $\chi^2$ added after minimization (``Pure statistics'') is shown as well as dotted curve, to illustrate the synergy.
}
\end{figure}

Let us first illustrate the complementarity and synergy between long-baseline experiments and PINGU with two examples as a function of $\deltacp$ in \figu{synergy}. In the left panel, it is clear that PINGU has a relatively flat performance in $\deltacp$, whereas the sensitivity of NO$\nu$A strongly depends on $\deltacp$. Therefore, the two experiment classes complement each other in terms of the parameter space. 
A similar example is shown in \figu{synergy}, right panel, where the two experiment classes clearly cover different parts of the parameter space. However, a $5 \sigma$ discovery for all values of $\deltacp$ can only come from the combination of the different experiments. It is noteworthy that in this case the two experiment classes are synergistic beyond a simple addition of statistics, see \Ref~\cite{Huber:2002rs} for a more detailed discussion. In order to illustrate that, we show the curve ``Pure statistics'', for which the $\chi^2$ of the two results was simply added after the individual parameter space minimization. Since, however, the parameter spaces are different, these minima are now allowed in a combined fit, and the sensitivity improves beyond a simple addition of statistics (compare curves ``Pure statistics'' and ``Combined'').

\begin{figure}[t]
\begin{center}
 \includegraphics[width=0.49\textwidth]{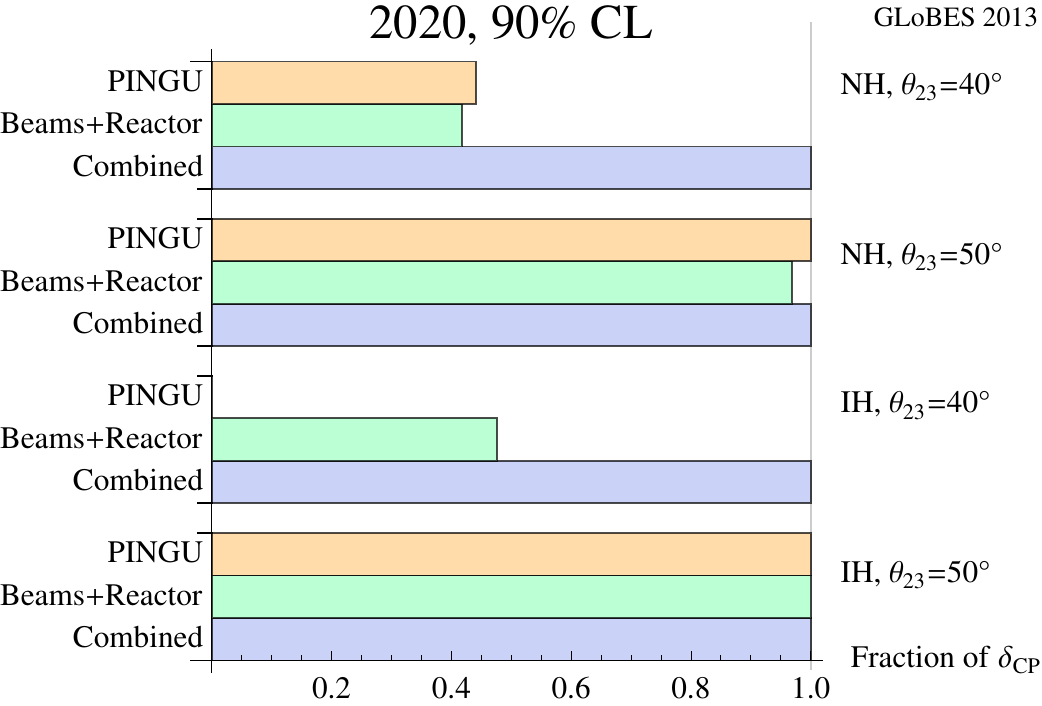} %
 \includegraphics[width=0.49\textwidth]{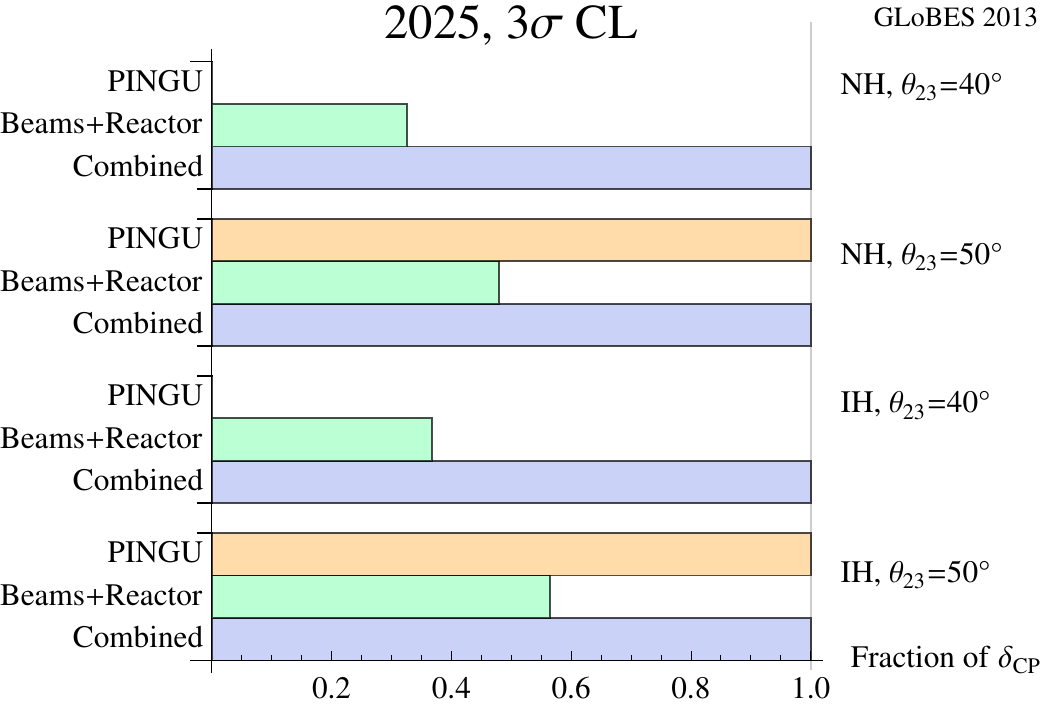}
\end{center}
\mycaption{\label{fig:sensallbeams} The fraction of $\deltacp$ for which the mass hierarchy can be discovered in 2020 (left panel, 90\% CL) or 2025 (right panel, $3\sigma$ CL). The different bar groups correspond to different (true) hierarchies and $\theta_{23}$, the individual bars to PINGU (three years, left panel and eight years, right panel, respectively), beams and reactor experiments (scenarios 2020 and 2025, respectively), and their combination.
}
\end{figure}

A more complete perspective on the parameter space is given  in \figu{sensallbeams}, where the fraction of $\deltacp$ for which the hierarchy can be discovered  is shown  in the scenarios 2020 (left panel, 90\% CL) and 2025 (right panel, $3\sigma$ CL). For $\theta_{23}=40^\circ$, the parameter space coverage in $\deltacp$ depends on the scenario and experiment class; for the normal hierarchy, PINGU seems to have a slight advantage, for the inverted hierarchy, beam and reactor experiments. For $\theta_{23}=50^\circ$, PINGU alone can reach 100\% coverage in $\deltacp$ in all cases. It is noteworthy that the combination of all experiments allows for a 90\% CL hint for the hierarchy in 2020 in all cases, and for a $3 \sigma$ discovery in 2025 --- in spite of considerably poorer individual performances especially for $\theta_{23} = 40^\circ$. Note however that the scenario 2025 relies on ProjectX.

\section{Neutrino beam to PINGU?}
\label{sec:beam}

\begin{figure}[t]
\begin{center}
 \includegraphics[width=0.7\textwidth]{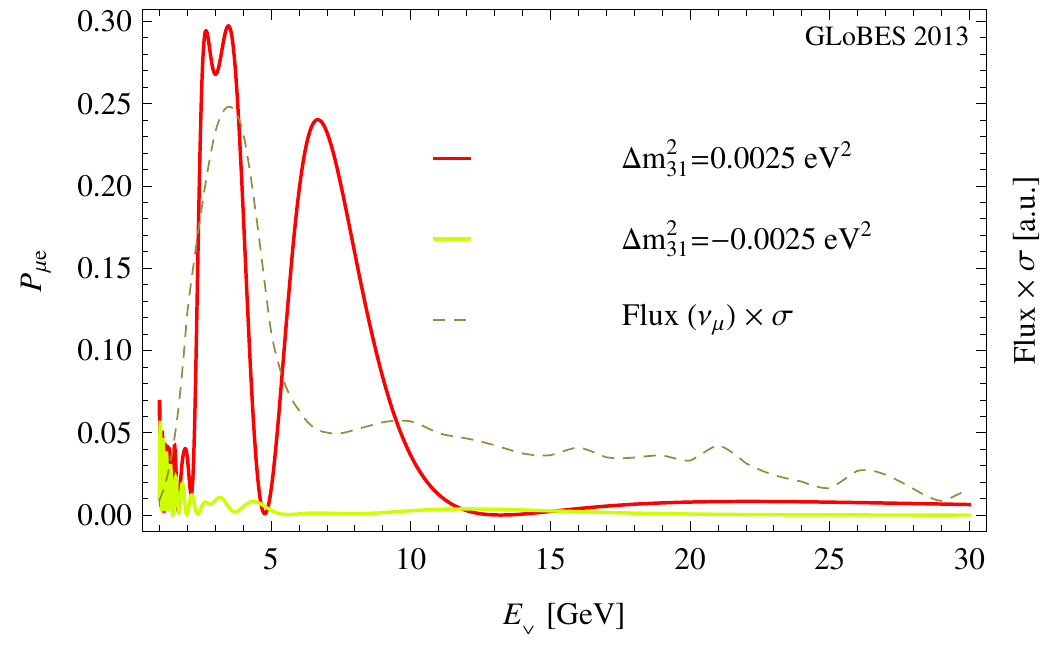}
\end{center}
\mycaption{\label{fig:oscplot} The $P_{\mu e}$ oscillation probability for a baseline $L = 11810 \, \mathrm{km}$ for $\ldm > 0$ and $\ldm < 0$ as a function of energy (for the other parameters, our standard values, $\theta_{23}=40^\circ$, and $\deltacp=0$ have been used). We also show the NuMI-like beam spectrum times cross section for a decay pipe length of 250~m for comparison (dashed curve). 
}
\end{figure}

In \Ref~\cite{Tang:2011wn} the requirements for a detector at the South Pole to receive a neutrino beam have been discussed, where it has been demonstrated that the mass hierarchy can be easily measured. After the discovery of a non-zero value of $\theta_{13}$, it was furthermore proposed that a low intensity beam from the Fermilab main injector would be sufficient for a high confidence mass hierarchy determination in spite of large irreducible backgrounds~\cite{Winter:LBNE}. Here we adopt this idea for the detector properties in this study.

We study the potential of  a setup with a neutrino beam from one of the major accelerator laboratories on the Northern hemisphere,  such as CERN, FNAL, JHF, DESY, or RAL. Since it turns out that the baselines from these laboratories to the South Pole are very similar and cross the Earth's outer core, we use $L \simeq 11810 \, \mathrm{km}$ corresponding to the baseline CERN-PINGU. The main sensitivity is expected to come from the $P_{\mu e}$ oscillation probability, since the beam produces mostly muon neutrinos. We therefore show   this  probability in  \figu{oscplot} for the normal and inverted mass hierarchies as a function of energy. It is clear that even by a total rate measurement of this channel the hierarchy can be measured. Using a beam spectrum from protons from the Fermilab main injector (the NuMI beam spectrum from the MINOS simulation in \Ref~\cite{Huber:2004ug} based on \Refs~\cite{Ables:1995wq,NUMIL714}), we observe that it peaks  between about 1~and 6~GeV. This is illustrated as dashed curve in \figu{oscplot}, where the product of flux and cross section is shown. In that energy range, the MSW resonance of the Earth's core and the parametric enhancement of the mantle-core-mantle profile~\cite{Akhmedov:1998ui,Akhmedov:1998xq,Petcov:1998su} lead to a strong enhancement of the oscillation probability, which is significantly larger than the vacuum oscillation amplitude $\propto \stheta$. This combination among parametric enhancement of the oscillation effect, a the beam spectrum peaking in that energy range, and a Megaton-size detector can be used for the mass hierarchy determination with, in fact, a very low intensity beam, such as from a short decay pipe.

Let us fix the boundary conditions similar to the NuMI beam: For the target power, we use 700~kW, as it is expected to be achieved within 2014 for the NO$\nu$A experiment, corresponding to $6.8 \, 10^{20} \, \mathrm{pot \, yr^{-1}}$, and for the total running time we use five years. Since the beam has to be tilted at an angle of about 68$^\circ$ to point towards the South Pole, a decay pipe as long as the one from the NuMI beam (675~m) is difficult to build. Therefore, we assume a shorter decay pipe of about 50 to 250~m only, at the expense of losing about one order of magnitude in flux.\footnote{The modification of the beam spectrum is estimated by assuming that the pion decays are exponentially distributed over the decay pipe length and that the neutrino takes about 25\% of the pion energy on average.} We use the $\nu_\mu$ running mode of the beam only, which will lead to a somewhat worse sensitivity for the inverted than normal hierarchy. 

For PINGU, we do not require directional resolution in that case, since the duty cycle of the beam can be used to suppress the backgrounds. We discuss the appearance channel $\nu_\mu \rightarrow \nu_e$ as the main source of information about the hierarchy, which means that electromagnetic cascade identification is, in principle, required. Since it is, however, difficult to distinguish electromagnetic cascades from hadronic or neutral current cascades, we make the most conservative assumption that all these cascades are treated as indistinguishable within one data sample, and that 30\% of the muon tracks are mis-identified as cascades.\footnote{This assumption corresponds to preliminary results for the ORCA detector at the spectral peak energy used here~\cite{Brunner:2013lua}.} Note that without flavor identification, the mass hierarchy can, of course, not be discriminated. However, if one can at least discriminate a part of the muon tracks from electromagnetic cascades on a statistical basis and make use of the fact that the other main background, the hadronic cascades, will be suppressed below the $\tau$ production threshold (where the beam spectrum peaks), the mass hierarchy determination becomes possible, see also discussions in \Refs~\cite{Brunner:2013lua,LujanPeschard:2013ud} for ORCA. Irreducible intrinsic beam $\nu_e$ and $\bar \nu_\mu$ backgrounds are included as well. For the energy resolution, we use $\Delta E/E=25\%$, which is the same as for the default setup above. Note, however, that for the signal event sample a better resolution may be expected, since the energy deposition in cascades is in fact easier to measure than in muon tracks. For this analysis, we include the T2K disappearance data, since the experiment itself cannot measure $\ldm$ as good as in the atmospheric case.

One of the critical factors for this measurement is how well the beam spectrum is known. We therefore test two different assumptions:
\begin{description}
 \item[$\text{NuMI}'_{50}$ $\rightarrow$ PINGU, ND] A very short decay pipe of 50~m is assumed, at the expense of the flux. However, it is assumed that one or more near detectors (ND) can be used to understand the beam spectrum and backgrounds at the level of 5\%. The analysis range spans $1$ to $30 \, \mathrm{GeV}$.
 \item[$\text{NuMI}'_{250}$ $\rightarrow$ PINGU, no ND] A considerably longer  decay pipe of 250~m is assumed. Since it is harder to deploy near detectors in this case, we assume that the total flux can be controlled at the level of 10\% without near detectors close to the peak flux. In addition, we assume that the analysis range covers the range between  $1$ and $10 \, \mathrm{GeV}$ only, where the flux is relatively large. The reason for this window is that it can be hardly assumed that, without ND, the high-E tail, where the flux is more than an order of magnitude lower than at the peak, is representative for the peak.\footnote{At high-E, there is sensitivity to the hierarchy which is more or less flat in energy, \ie, the systematics (which are here correlated among all bins) are important. This requires that one understands the spectral shape very well, since, for instance, a spectral tilt error would highly affect the sensitivity. We therefore only discuss the possibility of a large analysis range in the context of near detector(s).} 
\end{description}

\begin{figure}[t]
\begin{center}
 \includegraphics[width=0.48\textwidth]{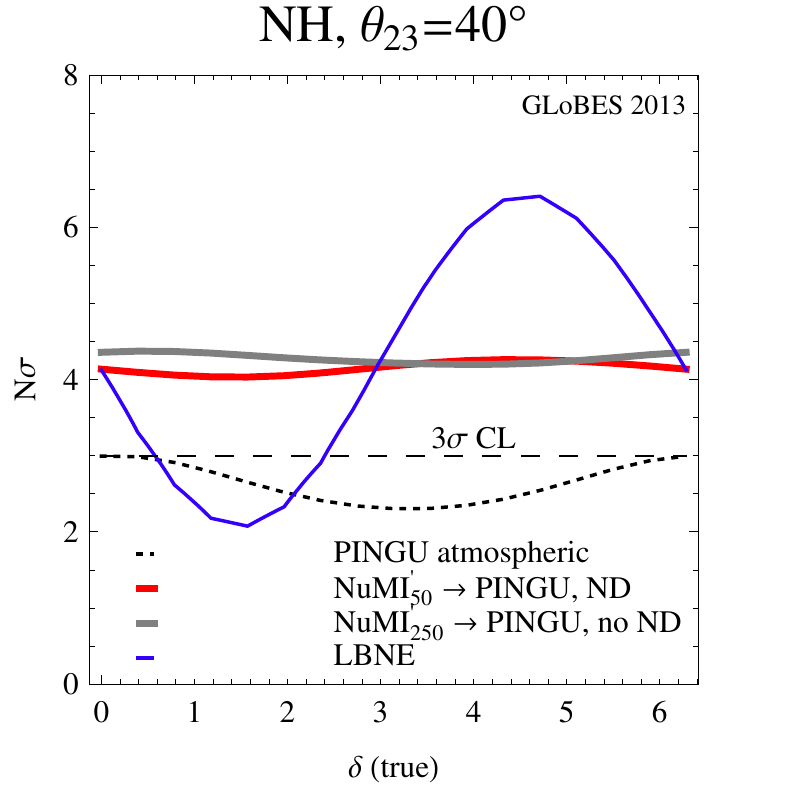} \includegraphics[width=0.48\textwidth]{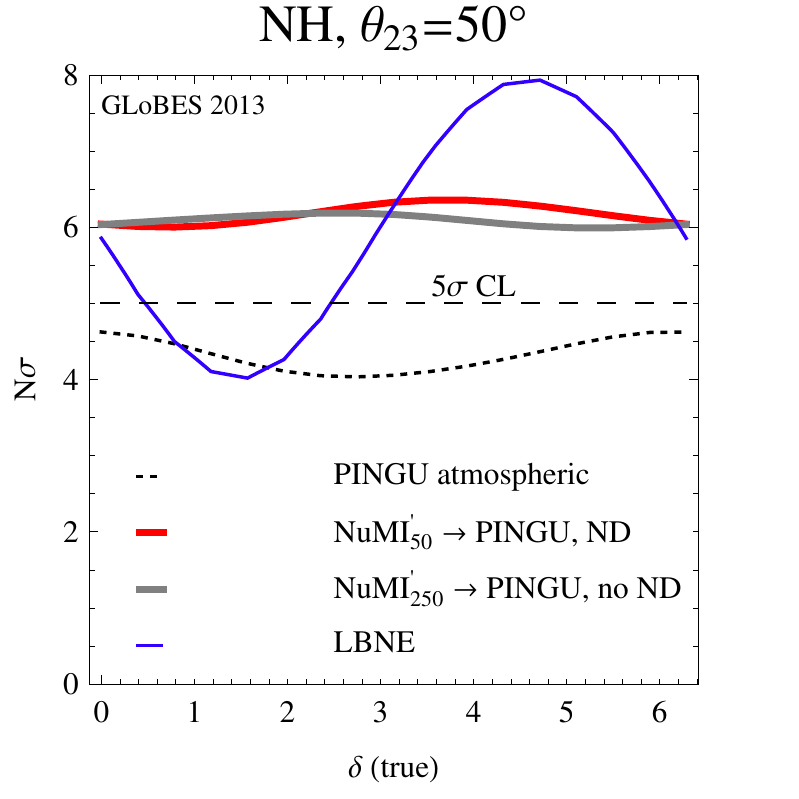} 
\end{center}
\mycaption{\label{fig:sensallbeam} The $\chi^2$ for the mass hierarchy discovery as a function of the true $\deltacp$ for the different  octants, as given in the plot captions. The dashed curves correspond to the default setup in \Tab~\ref{tab:sys} for eight years of PINGU operation using atmospheric neutrinos, the thick solid curves to the neutrino beam to PINGU with and without near detector(s) (ND) after five years of operation, and the thin solid curves to the LBNE experiment with a 10~kt liquid argon detector and five years of operation (2.5 years neutrinos and 2.5 years antineutrinos), see \Ref~\cite{Coloma:2012ji}. The beam to PINGU version with near detector uses a short 50~m decay pipe only and 5\% systematics, the one without near detector a 250~m decay pipe and 10\% systematics. In addition, the external information from ten years of T2K disappearance data is used for the beams.
}
\end{figure}

We show in \figu{sensallbeam} the sensitivity to the mass hierarchy comparing PINGU with atmospheric neutrinos and PINGU with a neutrino beam for the two different setups. The data taking times are different, since it is anticipated that such a beam experiment cannot be realized before 2020; therefore eight years of operation are assumed for the atmospheric neutrino setup. For comparison, we also show the LBNE experiment using five years of operation, for which, however, a later starting time of 2022 is already officially anticipated. One can clearly see that the PINGU beam experiment can determine the mass hierarchy almost everywhere in the parameter space at almost $4\sigma$ for $\theta_{23}=40^\circ$ and $6\sigma$ for $\theta_{23}=50^\circ$ irrespective of the conservative assumptions for the detector and beam. 
The versions including and excluding the near detector perform similar: one has better control of systematics, one has higher statistics. In fact, the sensitivities are similar to LBNE after the same running time, which has a somewhat stronger dependence on $\deltacp$. In that case, the detector is significantly smaller, but the shorter baseline, higher beam flux, and controlled systematics and flavor identification compensate for that. The results for the inverted hierarchy sensitivities are qualitatively similar at around $4\sigma$ irrespective of the octant. It is, however, noteworthy that in this case systematics control is somewhat more important than statistics, \ie, the option with near detector performs slightly better.

Finally, note that optimizing the target geometry and decay tunnel length may lead to significantly improved sensitivities of the beam experiment; it is therefore not clear how effective this option  for the hierarchy sensitivity is. Furthermore, it may be used as a backup if $\theta_{23}$ is indeed small or the experiment properties, which are partially not needed for the beam (such as directional resolution), are overestimated (\cf, conservative setup curves in \figu{sensall}).

\section{Summary and conclusions}

We have discussed the mass hierarchy sensitivity of PINGU using cosmic ray interactions in the Earth's atmosphere  and a neutrino beam as neutrino sources, and compared it to the one of existing long-baseline and reactor experiments on the same timescale, driven by the physics of NO$\nu$A. We have in all cases included the full three flavor oscillation framework with the full parameter correlations and degeneracies, as it is state-of-the-art for any beam analysis, by directly simulating PINGU with GLoBES. In addition, we have considered an unprecedentedly  large number of systematical errors for the atmospheric analysis.

For the mass hierarchy determination with atmospheric neutrinos, we have shown that conventional estimators for the degenerate solution do not work everywhere in parameter space, which means that correlations other than with $\ldm$ can be important.  In order not to mis-interpret any tension with external data as hierarchy sensitivity, we have tested PINGU in combination with 10 years of T2K disappearance data in a self-consistent way -- with a marginal impact on the hierarchy sensitivity. In fact,  we have shown that PINGU alone can measure $\ldm$ with an error of about 0.9\% ($1\sigma$), which means that PINGU may provide the best measurement of $\ldm$ at that time, and that we have avoided the combination with external data on the atmospheric parameters.

The sensitivity also strongly depends on the true values of the parameters, such as $\deltacp$, the $\theta_{23}$ octant, and the hierarchy. As far as the experiment properties are concerned, exposure, directional resolution, energy resolution, energy threshold, and cascade mis-identification have been found to be the driving impact factors to be optimized for. On the other hand, taking into account the full sky coverage and energies up to 50~GeV, a number of systematical errors related to the atmospheric neutrino flux, cross sections, and matter density uncertainties can be well controlled. The measurement will therefore be very robust with respect to systematical errors. 

Taking into account all factors, a 90\%CL to $4\sigma$ discovery after three years of operation seems conceivable including the full parameter degeneracy for the atmospheric neutrino analysis, depending on the true value of $\theta_{23}$ and the exploitation of the potential for experimental optimization. The sensitivity is complementary in parameter space to that of existing beam and reactor experiments, driven by the NO$\nu$A experiment, and in some cases synergistic effects are present which go beyond the simple addition of statistics if the experiments are combined because of different parameter space topologies. For instance, a $3 \sigma$ discovery in 2025 everywhere in the parameter space is possible if the NuMI beam is upgraded by ProjectX even with the default setup for PINGU used in this study. 

Finally, we have discussed the option to send a beam from any of the major accelerator laboratories on the Northern hemisphere to the South Pole, where the combination among parametric enhancement from the mantle-core-mantle profile of the Earth, beam spectrum peaking in the right energy range, and Megaton-sized detector in that range offers a unique physics potential. At the example of a NuMI-like beam with a significantly shortened decay pipe (50~m including a near detector or 250~m without near detector), we have demonstrated that a $4\sigma$ to $6\sigma$ discovery after five years of operation could be feasible, and the sensitivity is comparable to the LBNE experiment. This option may be an interesting backup if the true $\theta_{23}$ is indeed small or the experiment properties, such as directional resolution, turn out to be overestimated. One advantage of this option is that the source location does not yet needed to be specified -- such a beam could come from FNAL, CERN, JHF, DESY, or RAL. The actual technical implementation requires, however, further study. Maybe even CP violation studies become possible if the detector can be further upgraded, a concept known as MICA (``Megaton Ice Cherenkov Array''), for which a fiducial masses of order 10~Mt may be reachable already below 1~GeV~\cite{Boser:2013oaa}.

In conclusion, we have demonstrated that PINGU can measure the mass hierarchy using atmospheric neutrinos using the same assumptions and machinery which has been used for the beam experiments in the past. However, the final confidence will depend on the experiment parameters and true oscillation parameter values, and a high confidence level determination may require a significant running time of the order of a few years if the full parameter space degeneracy is included. 

\subsubsection*{Acknowledgments}

I would like to thank M. Day, J. Leute, U. Katz, M. Kowalski, X. Qian, E. Resconi, and the members of the PINGU collaboration for useful discussions related to the subjects in this study.

This work has been supported by DFG grants WI 2639/3-1 and WI 2639/4-1, the FP7 Invisibles network (Marie Curie
Actions, PITN-GA-2011-289442), and the ``Helmholtz Alliance for Astroparticle Physics HAP'', funded by the Initiative and Networking fund of the Helmholtz association.


\begin{thebibliography}{10}

\bibitem{Abe:2011fz}
DOUBLE-CHOOZ Collaboration, Y.~Abe {\em et~al.},
\newblock Phys.Rev.Lett. {\bf 108}, 131801 (2012), arXiv:1112.6353.

\bibitem{An:2012eh}
DAYA-BAY Collaboration, F.~An {\em et~al.},
\newblock Phys.Rev.Lett. {\bf 108}, 171803 (2012), arXiv:1203.1669.

\bibitem{Ahn:2012nd}
RENO collaboration, J.~Ahn {\em et~al.},
\newblock Phys.Rev.Lett. {\bf 108}, 191802 (2012), arXiv:1204.0626.

\bibitem{GonzalezGarcia:2012sz}
M.~Gonzalez-Garcia, M.~Maltoni, J.~Salvado, and T.~Schwetz,
\newblock JHEP {\bf 1212}, 123 (2012), arXiv:1209.3023.

\bibitem{Fogli:2012ua}
G.~Fogli {\em et~al.},
\newblock Phys.Rev. {\bf D86}, 013012 (2012), arXiv:1205.5254.

\bibitem{Tortola:2012te}
D.~Forero, M.~Tortola, and J.~Valle,
\newblock Phys.Rev. {\bf D86}, 073012 (2012), arXiv:1205.4018.

\bibitem{Huber:2009cw}
P.~Huber, M.~Lindner, T.~Schwetz, and W.~Winter,
\newblock JHEP {\bf 0911}, 044 (2009), arXiv:0907.1896.

\bibitem{Itow:2001ee}
Y.~Itow {\em et~al.},
\newblock Nucl. Phys. Proc. Suppl. {\bf 111}, 146 (2001), hep-ex/0106019.

\bibitem{Ayres:2004js}
NOvA Collaboration, D.~Ayres {\em et~al.},
\newblock (2004), arXiv:hep-ex/0503053.

\bibitem{Coloma:2012ji}
P.~Coloma, P.~Huber, J.~Kopp, and W.~Winter,
\newblock Phys. Rev. {\bf D87}, 033004 (2013), arXiv:1209.5973.

\bibitem{Albright:2006cw}
C.~H. Albright and M.-C. Chen,
\newblock Phys. Rev. {\bf D74}, 113006 (2006), arXiv:hep-ph/0608137.

\bibitem{Wolfenstein:1978ue}
L.~Wolfenstein,
\newblock Phys. Rev. {\bf D17}, 2369 (1978).

\bibitem{Mikheev:1985gs}
S.~P. Mikheev and A.~Y. Smirnov,
\newblock Sov. J. Nucl. Phys. {\bf 42}, 913 (1985).

\bibitem{Mikheev:1986wj}
S.~P. Mikheev and A.~Y. Smirnov,
\newblock Nuovo Cim. {\bf C9}, 17 (1986).

\bibitem{Barger:2006vy}
V.~Barger {\em et~al.},
\newblock Phys.Rev. {\bf D74}, 073004 (2006), arXiv:hep-ph/0607177.

\bibitem{Barger:2007jq}
V.~Barger, P.~Huber, D.~Marfatia, and W.~Winter,
\newblock Phys. Rev. {\bf D76}, 053005 (2007), arXiv:hep-ph/0703029.

\bibitem{Samanta:2006sj}
A.~Samanta,
\newblock Phys.Lett. {\bf B673}, 37 (2009), arXiv:hep-ph/0610196.

\bibitem{Ghosh:2012px}
A.~Ghosh, T.~Thakore, and S.~Choubey,
\newblock JHEP {\bf 1304}, 009 (2013), arXiv:1212.1305.

\bibitem{Blennow:2012gj}
M.~Blennow and T.~Schwetz,
\newblock JHEP {\bf 1208}, 058 (2012), arXiv:1203.3388.

\bibitem{Barger:2012fx}
V.~Barger {\em et~al.},
\newblock Phys.Rev.Lett. {\bf 109}, 091801 (2012), arXiv:1203.6012.

\bibitem{Li:2013zyd}
Y.-F. Li, J.~Cao, Y.~Wang, and L.~Zhan,
\newblock (2013), arXiv:1303.6733.

\bibitem{Oyama:2012tq}
Y.~Oyama, A.~Shimizu, and K.~Kohri,
\newblock Phys.Lett. {\bf B718}, 1186 (2013), arXiv:1205.5223.

\bibitem{Koskinen:2011zz}
D.~J. Koskinen,
\newblock Mod.Phys.Lett. {\bf A26}, 2899 (2011).

\bibitem{Clark:2012hya}
IceCube/PINGU Collaboration, K.~Clark and D.~Cowen,
\newblock Nucl.Phys.Proc.Suppl. {\bf 233}, 223 (2012).

\bibitem{ORCA}
U.~Katz,
\newblock {The ORCA Option for KM3NeT},
\newblock {Talk given at the XV International Workshop on Neutrino Telescopes,
  Venedig, Italy}.

\bibitem{Akhmedov:2012ah}
E.~K. Akhmedov, S.~Razzaque, and A.~Y. Smirnov,
\newblock JHEP {\bf 02}, 082 (2013), arXiv:1205.7071.

\bibitem{Agarwalla:2012uj}
S.~K. Agarwalla, T.~Li, O.~Mena, and S.~Palomares-Ruiz,
\newblock (2012), arXiv:1212.2238.

\bibitem{Franco:2013in}
D.~Franco {\em et~al.},
\newblock JHEP {\bf 1304}, 008 (2013), arXiv:1301.4332.

\bibitem{Ohlsson:2013epa}
T.~Ohlsson, H.~Zhang, and S.~Zhou,
\newblock (2013), arXiv:1303.6130.

\bibitem{Esmaili:2013fva}
A.~Esmaili and A.~Y. Smirnov,
\newblock (2013), arXiv:1304.1042.

\bibitem{Ribordy:2013xea}
M.~Ribordy and A.~Y. Smirnov,
\newblock (2013), arXiv:1303.0758.

\bibitem{Dick:2000fn}
K.~Dick, M.~Freund, P.~Huber, and M.~Lindner,
\newblock Nucl. Phys. {\bf B588}, 101 (2000), hep-ph/0006090.

\bibitem{Winter:2005we}
W.~Winter,
\newblock Phys. Rev. {\bf D72}, 037302 (2005), hep-ph/0502097.

\bibitem{Fargion:2010vb}
D.~Fargion, D.~D'Armiento, P.~Desiati, and P.~Paggi,
\newblock Astrophys.J. {\bf 758}, 3 (2012), arXiv:1012.3245.

\bibitem{Tang:2011wn}
J.~Tang and W.~Winter,
\newblock JHEP {\bf 1202}, 028 (2012), arXiv:1110.5908.

\bibitem{Huber:2004ka}
P.~Huber, M.~Lindner, and W.~Winter,
\newblock Comput. Phys. Commun. {\bf 167}, 195 (2005), hep-ph/0407333,
\newblock {\tt http://www.mpi-hd.mpg.de/lin/globes/}.

\bibitem{Huber:2007ji}
P.~Huber, J.~Kopp, M.~Lindner, M.~Rolinec, and W.~Winter,
\newblock Comput. Phys. Commun. {\bf 177}, 432 (2007), hep-ph/0701187.

\bibitem{Athar:2012it}
M.~Sajjad~Athar, M.~Honda, T.~Kajita, K.~Kasahara, and S.~Midorikawa,
\newblock Phys.Lett. {\bf B718}, 1375 (2013), arXiv:1210.5154.

\bibitem{Honda:2006qj}
M.~Honda, T.~Kajita, K.~Kasahara, S.~Midorikawa, and T.~Sanuki,
\newblock Phys.Rev. {\bf D75}, 043006 (2007), arXiv:astro-ph/0611418.

\bibitem{PINGU}
{PINGU collaboration},
\newblock official plots.

\bibitem{Huber:2005jk}
P.~Huber, M.~Lindner, M.~Rolinec, and W.~Winter,
\newblock Phys. Rev. {\bf D73}, 053002 (2006), hep-ph/0506237.

\bibitem{Winter:2008cn}
W.~Winter,
\newblock Phys. Rev. {\bf D78}, 037101 (2008), arXiv:0804.4000.

\bibitem{Geller:2001ix}
R.~J. Geller and T.~Hara,
\newblock Phys. Rev. Lett. {\bf 49}, 98 (2001), hep-ph/0111342.

\bibitem{Ciuffoli:2013rza}
E.~Ciuffoli, J.~Evslin, and X.~Zhang,
\newblock (2013), arXiv:1305.5150.

\bibitem{deGouvea:2005hk}
A.~de~Gouvea, J.~Jenkins, and B.~Kayser,
\newblock Phys. Rev. {\bf D71}, 113009 (2005), hep-ph/0503079.

\bibitem{Nunokawa:2005nx}
H.~Nunokawa, S.~J. Parke, and R.~Zukanovich~Funchal,
\newblock Phys.Rev. {\bf D72}, 013009 (2005), arXiv:hep-ph/0503283.

\bibitem{deGouvea:2005mi}
A.~de~Gouvea and W.~Winter,
\newblock Phys. Rev. {\bf D73}, 033003 (2006), arXiv:hep-ph/0509359.

\bibitem{Huber:2002mx}
P.~Huber, M.~Lindner, and W.~Winter,
\newblock Nucl. Phys. {\bf B645}, 3 (2002), hep-ph/0204352.

\bibitem{BackhouseNOVA}
C.~Backhouse,
\newblock {Status of NO$\nu$A},
\newblock Talk given at NNN2012.

\bibitem{Huber:2002rs}
P.~Huber, M.~Lindner, and W.~Winter,
\newblock Nucl. Phys. {\bf B654}, 3 (2003), hep-ph/0211300.

\bibitem{Winter:LBNE}
W.~Winter,
\newblock {Superbeam FNAL-PINGU?},
\newblock {Analysis and slides presented within the LBNE reconfiguration
  effort, May 2012. See also talk at TeVPA 2012.}

\bibitem{Huber:2004ug}
P.~Huber, M.~Lindner, M.~Rolinec, T.~Schwetz, and W.~Winter,
\newblock Phys. Rev. {\bf D70}, 073014 (2004), arXiv:hep-ph/0403068.

\bibitem{Ables:1995wq}
MINOS, E.~Ables {\em et~al.},
\newblock FERMILAB-PROPOSAL-0875.

\bibitem{NUMIL714}
M.~Diwan, M.~Messier, B.~Viren, and L.~Wai,
\newblock A study of {$\nu_\mu \rightarrow \nu_e$} sensitivity in minos,
\newblock NUMI-L-714 (2001).

\bibitem{Akhmedov:1998ui}
E.~K. Akhmedov,
\newblock Nucl.Phys. {\bf B538}, 25 (1999), arXiv:hep-ph/9805272.

\bibitem{Akhmedov:1998xq}
E.~K. Akhmedov, A.~Dighe, P.~Lipari, and A.~Smirnov,
\newblock Nucl.Phys. {\bf B542}, 3 (1999), arXiv:hep-ph/9808270.

\bibitem{Petcov:1998su}
S.~Petcov,
\newblock Phys.Lett. {\bf B434}, 321 (1998), arXiv:hep-ph/9805262.

\bibitem{Brunner:2013lua}
J.~Brunner,
\newblock (2013), arXiv:1304.6230.

\bibitem{LujanPeschard:2013ud}
C.~Lujan-Peschard, G.~Pagliaroli, and F.~Vissani,
\newblock (2013), arXiv:1301.4577.

\bibitem{Boser:2013oaa}
S.~B{\"o}ser, M.~Kowalski, L.~Schulte, N.~L. Strotjohann, and M.~Voge,
\newblock (2013), arXiv:1304.2553.

\end{thebibliography}

\end{document}